\documentclass[preprintnumbers,article,amsmath,amssymb,floatfix,10pt,prd,onecolumn,
superscriptaddress,nofootinbib]{revtex4}
\usepackage[colorlinks=true, pdfstartview=FitV, linkcolor=blue, citecolor=red, urlcolor=magenta]{hyperref}
\usepackage{bbm}
\usepackage{amsfonts}
\usepackage{mathrsfs}
\usepackage{latexsym}
\usepackage{epsfig}
\usepackage{epstopdf}
\usepackage{epstopdf}
\usepackage{graphicx}
\usepackage{amssymb}
\usepackage{amsmath}
\usepackage{dcolumn}
\usepackage{bm}
\usepackage{float}
\usepackage{color}
\usepackage{comment}
\usepackage{xcolor}
\begin{document}
\title{Thermal analysis and Joule-Thomson expansion of black hole exhibiting metric-affine gravity}
\author{Muhammad Yasir}
\email{yasirciitsahiwal@gmail.com}\affiliation{Department of Mathematics, Shanghai University,
Shanghai, 200444, Shanghai, People's Republic of China}
\author{Xia Tiecheng}
\email{xiatc@shu.edu.cn}\affiliation{Department of Mathematics, Shanghai University,
Shanghai, 200444, Shanghai, People's Republic of China}
\author{Faisal Javed}
\email{faisaljaved.math@gmail.com}\affiliation{Department of
Physics, Zhejiang Normal University, Jinhua 321004, People's
Republic of China}

\affiliation{Zhejiang Institute of
Photoelectronics and Zhejiang Institute for Advanced Light Source,
Zhejiang Normal University, Jinhua, Zhejiang 321004, China}

\author{G. Mustafa}
\email{gmustafa3828@gmail.com}\affiliation{Department of Physics, Zhejiang Normal University, Jinhua 321004, People's
Republic of China}

\affiliation{New Uzbekistan University, Mustaqillik Ave. 54, Tashkent 100007, Uzbekistan.}

\begin{abstract}
This study examines a recently hypothesized black hole, which is a
perfect solution of metric-affine gravity with a positive
cosmological constant, and its thermodynamic features as well as the
Joule-Thomson expansion. We develop some thermodynamical quantities,
such as volume, Gibbs free energy, and heat capacity, using the
entropy and Hawking temperature. We also examine the first law of
thermodynamics and thermal fluctuations, which might eliminate
certain black hole instabilities. In this regard, a phase transition
from unstable to stable is conceivable when the first law order
corrections are present. Besides that, we study the efficiency of
this system as a heat engine and the effect of metric-affine gravity
for physical parameters $q_e$, $q_m$, $\kappa_{\mathrm{s}}$,
$\kappa_{\mathrm{d}}$ and $\kappa_{\mathrm{sh}}$. Further, we study
the Joule-Thomson coefficient, and the inversion temperature and
also observed the isenthalpic curves in the $T_i -P_i$ plane. In
metric-affine gravity, a comparison is made between the Van der
Waals fluid and the black hole to study their similarities and
differences.
\end{abstract}
\date{\today}
\maketitle
\section{Introduction}

One of the most attractive and challenging subjects of study is the
geometrical structure of black hole (BH) in general relativity (GR)
and modified theories of gravity \cite{1}. The thermal
characteristics of BHs and their behavior are analyzed by well-known
four laws of BH mechanics \cite{2,3}. After that the work of
Bekenstein, first-time Stephen Hawking presented the existence of BH
radiations and formalized the tunneling process very near the BH
horizon due to the vacuum fluctuations. It is observed that the
small amount of heat quantity leads to the eccentricity of quantum
mechanics \cite{4,5}. In the literature \cite{7}, it is noted that
the BHs contain thermodynamic features like temperature and entropy.
At the BH horizon, the Hawking temperature is proportional to its
surface gravity due to BH behaves like a thermodynamical system. It
is certified that results of \cite{8} are useful to all classical
BHs at thermodynamic equilibrium.

The Hawking temperature phase transition takes place after the
justification of a phase structure isomorphic associated with the
Van der Waals liquid-gas system in Kerr RN-AdS BH \cite{9} and
RN-AdS BH \cite{9,10}.  Till then, in all BH thermodynamic studies,
mass, volume, and pressure, the crucial thermodynamic variables were
missing. The foundation of pressure to this field is completed
through cosmological constant, which also has other basic
implications such as the consistency of Smarr's relation with the
first law \cite{11}. The cosmological constant ($\Lambda$) is taken
as the thermodynamic pressure and the respective first law of
thermodynamics was simultaneously modified by the expansion of phase
space with a $PdV$ term, leading to a novel understanding of the BH
mass \cite{12}. The new perspective on mass with the cosmological
constant in BH thermodynamics formulated phenomenal consequences in
classical thermodynamics. Kubiznak et al. \cite{13,14} presented AdS
BH as a van der Waals system and investigated the critical behavior
of BH through $P-v$ isotherm, Gibbs free energy, critical exponents
and coincidence curves, which are all presented to be similar to the van
der Waals case. Moreover, these similar features were obtained on
various AdS BHs in Refs. \cite{15,16,17,18,19,20}. Similarities to
classical thermodynamics like holographic heat engines \cite{21},
Joule Thomson expansion, phase transitions and
Clausius-Clapeyron equation are also studied in \cite{23,24}.  Javed et al. \cite{aa1}  investigated the thermodynamics of charged and uncharged BHs in symmetric teleparallel gravity. They also studied the thermal fluctuations and phase transition of considered BHs. The dynamical configurations of thin-shell developed from BHs in metric affine gravity composed with scalar field are studied in \cite{gt1}. Some interesting physical characteristics of various BH solutions are dicussed in \cite{p1}-\cite{p3}. 

In addition, Joule-Thomson expansion was
investigated in AdS BHs by {\"O}kc{\"u} and Aydiner \cite{22},
further it proceeds to the isenthalpic process by which gas expands
through a porous plug from a high-pressure section to a low-pressure
section. The researchers also analyzed the Joule-Thomson expansion phenomenon in Kerr-AdS black holes within the extended phase space \cite{22a}. They examined both isenthalpic and numerical inversion curves in the temperature-pressure plane, illustrating regions of cooling and heating for Kerr-AdS black holes. Additionally, they computed the ratio between the minimum inversion temperature and critical temperature for Kerr-AdS black holes \cite{22a}. This pioneering work was
generalized to quintessence holographic superfluids of RN BHs in
$f(R)$ gravity \cite{26,27,28}. More recently, we studied in detail
the consequence of the dimensionality on the Joule-Thomson expansion
in Ref. \cite{28}. It was presented that in \cite{28,29}, the
ratio between critical temperature decreases and minimum inversion
temperature with the dimensionality $d$ while it retrieves the
results when $d=4$. In this paper, we investigate the existence of metric-affine gravity
should influence the Joule-Thomson expansion, which also is
motivated by the progress in our understanding of metric-affine gravity. Here, we present that the chosen strategy is contextualized
not only for the BH in Metric-Affine gravity but also for those in
other alternative theories of gravity where new gravitational modes
well developed.

This paper is devoted to explore the effects of metric-affine gravity on  the phase transition  of BH geometry and also study the Joule-Thomson expansion. The formation of the current paper is as
written. In Sec. II, we study a brief review of our new class of BH in metric-affine
gravity. In  Sec. III, we formulate the thermodynamic quantities
like temperature, pressure, Gibbs free energy, and heat engine.
Next, in Section IV, we introduce a Joule-Thomson expansion for a
classical physical quantity. Finally, we present a few closing
remarks.

\section{A Brief Review on Black Hole in Metric-Affine Gravity}

General relativity is the most successful and physically acceptable
theory of gravity that precisely describe the gravitational
interaction among the space-time geometry and the characteristics of
matter via energy-momentum tensor. From a geometrical perspective,
the Lorentzian metric tensor $g_{\mu \nu}$ is considered to study
the smooth manifold which is used to develop the Levi-Civita affine
connection $\Gamma^\lambda _{\mu \nu}$. To establish a model where
the largest family of BH solutions with dynamical torsion and
nonmetricity in metric affine gravity can be found, a propagating
traceless nonmetricity tensor must be taken into account in the
gravitational action of metric affine gravity. As a
geometrical correction to GR, a quadratic parity-preserving action
presenting a dynamical traceless nonmetricity tensor in this
situation given as  \cite{y1,y2,y3,y5,y5a}:
\begin{widetext}
\begin{equation}\label{1}
S=\int d^4 x \sqrt{-g}\left\{\mathcal{L}_{\mathrm{m}}+\frac{1}{16
\pi}\left[-R+2 f_1 \tilde{R}_{(\lambda \rho) \mu \nu}
\tilde{R}^{(\lambda \rho) \mu \nu}+2 f_2\left(\tilde{R}_{(\mu
\nu)}-\hat{R}_{(\mu \nu)}\right)\left(\tilde{R}^{(\mu
\nu)}-\hat{R}^{(\mu \nu)}\right)\right]\right\},
\end{equation}
\end{widetext}
where $ \tilde{R}^{(\lambda \rho) \mu \nu}$ and $\tilde{R}_{(\mu\nu)}$ are the affine-connected form of Riemann and Ricci tensors. Here, $R$ denotes the Ricci scalar, $g$ is determinant of metric tensor, $\mathcal{L}_{\mathrm{m}}$ depicts the matter Lagrangian and $f_1$, $f_2$ are Lagrangian coefficients. This solution can also be easily generalized to take into account the cosmological constant and Coulomb electromagnetic fields with electric charge ($q_e$) and magnetic charge ($q_m$), which are decoupled from torsion \cite{y5a1,y5a2}. This is supposing the minimal coupling principle.
\begin{widetext}
\begin{eqnarray}
\begin{aligned}\label{2}
\tilde{R}^{(\lambda \rho)}_{\mu \nu} & =\frac{1}{2} T_{\mu \nu}^\sigma Q_\sigma^{\lambda \rho}+\tilde{\nabla}_{[\nu} Q_{\mu]}^{\lambda \rho}, \\
\tilde{R}_{(\mu \nu)}-\hat{R}_{(\mu \nu)} & =\tilde{\nabla}_{(\mu} Q_{\nu) \lambda}^\lambda+Q_{\lambda \rho(\mu} Q_{\nu)}^{\lambda \rho}-\tilde{\nabla}_\lambda Q_{(\mu \nu)}^\lambda-Q^{\lambda \rho}{ }_\lambda Q_{(\mu \nu) \rho}+T_{\lambda \rho(\mu} Q_{\nu \rho}^{\lambda \rho},
\end{aligned}
\end{eqnarray}
\end{widetext}
these variations represent the third Bianchi of GR.  By executing changes of above  equations  with respect to the co-frame field and the anholonomic interrelation, the following field equations are established  $Y 1_\mu^\nu =8 \pi \theta_\mu^\nu$ and $Y 2^{\lambda \mu \nu} =4 \pi \Delta^{\lambda \mu \nu}$, where $Y 1_\mu{ }^\nu$ and $Y 2^{\lambda \mu \nu}$ are tensor quantities. Utilizing $\Delta^{\lambda \mu \nu}$ and $\theta_\mu{ }^\nu$  to study the hyper momentum density  and canonical energy-momentum tensors of matter, expressed as
\begin{eqnarray} \label{3}
\begin{aligned}
\Delta^{\lambda \mu \nu} & =\frac{e^{a \lambda} e_b \mu}{\sqrt{-g}} \frac{\delta\left(\mathcal{L}_m \sqrt{-g}\right)}{\delta \omega^a{ }_{b \nu}},\\
\theta_\mu^\nu & =\frac{e^a{ }_\mu}{\sqrt{-g}} \frac{\delta\left(\mathcal{L}_m \sqrt{-g}\right)}{\delta e^a \nu}.
\end{aligned}
\end{eqnarray}
Therefore, both matter represents act as sources of the extended gravitational field. In this scenario metric-affine geometries utilizing  the Lie algebra of the general linear group GL(4, R) in  anholonomic interrelation. This hypermomentum present its proper decomposition into shear, spin and dilation  currents \cite{y5, y5a}. Furthermore, the effective gravitational action of the model provided in terms of these properties. The parameterizations of the spherically symmetric static spacetime is \cite{ y5a2,fg2,y5a3,y5a4,y5a5}
\begin{equation}\label{4}
ds^2=-\Psi(r)dt^2+\Psi^{-1}(r)dr^2 +r^2d\theta^2+r^2\sin^2\theta
d\phi^2.
\end{equation}
comparison with the standard case of GR , in emission process, a matter currents coupled to torsion and nonmetricity in general splitting of the energy levels will potentially affect this spectrum and efficiency. Interestingly, the performance of a perturbative interpretation on energy-momentum tensor  in vacuum fluctuations of quantum field coupled to the torsion as well as nonmetricity tensors of the solution, in order to study the rate of dissipation obtain on its event horizon, which would also cover the  further corrections with respect to the system of GR \cite{y5a6, y5a7}. The metric function (Reissner-Nordstr¨om-de Sitter-like) is defined as \cite{y5a}
\begin{widetext}
\begin{equation}\label{6}
\Psi(r)=1-\frac{2 m}{r}+\frac{d_1 \kappa_{\mathrm{s}}^2-4 e_1
\kappa_{\mathrm{d}}^2-2 f_1
\kappa_{\mathrm{sh}}^2+q_{\mathrm{e}}^2+q_{\mathrm{m}}^2}{r^2}+\frac{\Lambda}{3}
r^2,
\end{equation}
\end{widetext}
which represents the broadest family-charged BH models obtained in
metric affine gravity with real constants $e_1$ and $d_1$. Here,
$\kappa_{\mathrm{sh}}$, $\kappa_{\mathrm{s}}$ and
$\kappa_{\mathrm{d}}$ represent the shear, spin and dilation
charges, respectively.

\section{Thermodynamics}

\begin{figure}
\begin{minipage}{14pc}
\includegraphics[width=16pc]{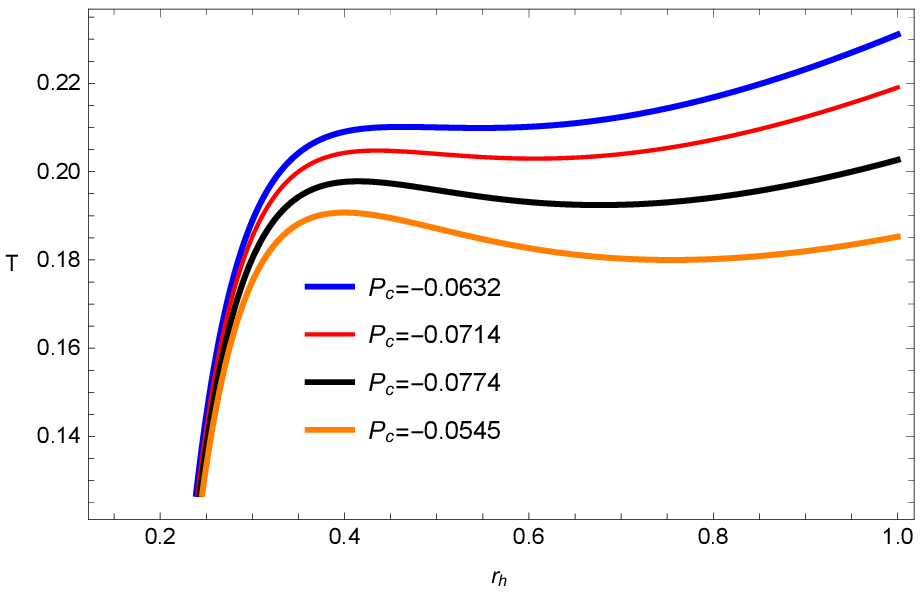}
\caption{\label{f1} Plot of temperature $T$ with fixed $q_{e}=0.28$;
$q_m=0.08$; $d_1=0.004$; $f_1=0.313$; $\kappa _d =0.02$; $\kappa_s
=0.8$ and $e_{1}=0.4$.}
\end{minipage}\hspace{3pc}%
\begin{minipage}{14pc}
\includegraphics[width=16pc]{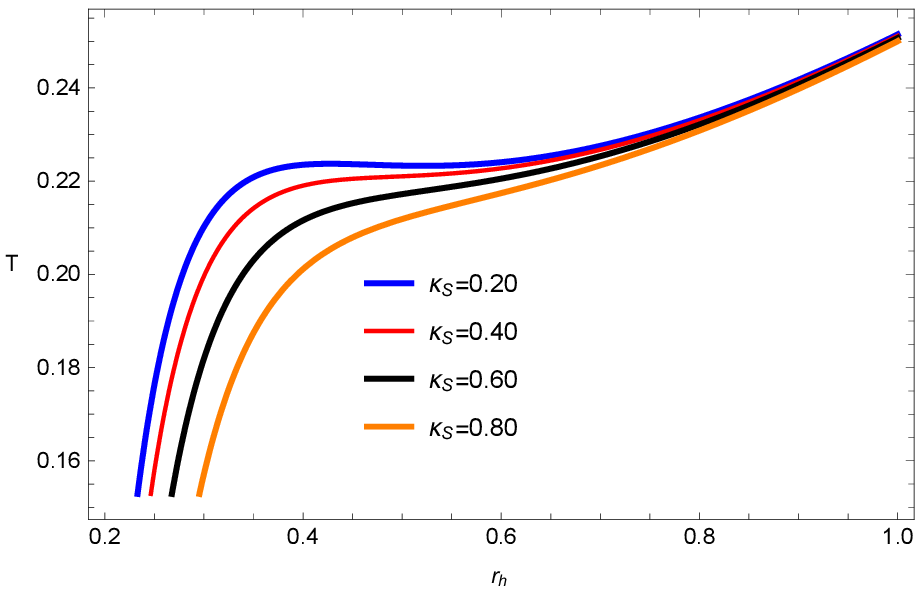}
\caption{\label{f2} Plot of temperature $T$ with fixed $q_{e
}=0.28$; $q_m=0.08$; $d_1=0.004$; $f_1=0.313$; $\kappa_d =0.02$ and
$e_1=0.4$.}
\end{minipage}\hspace{3pc}%
\end{figure}

\begin{figure}
\begin{minipage}{14pc}
\includegraphics[width=16pc]{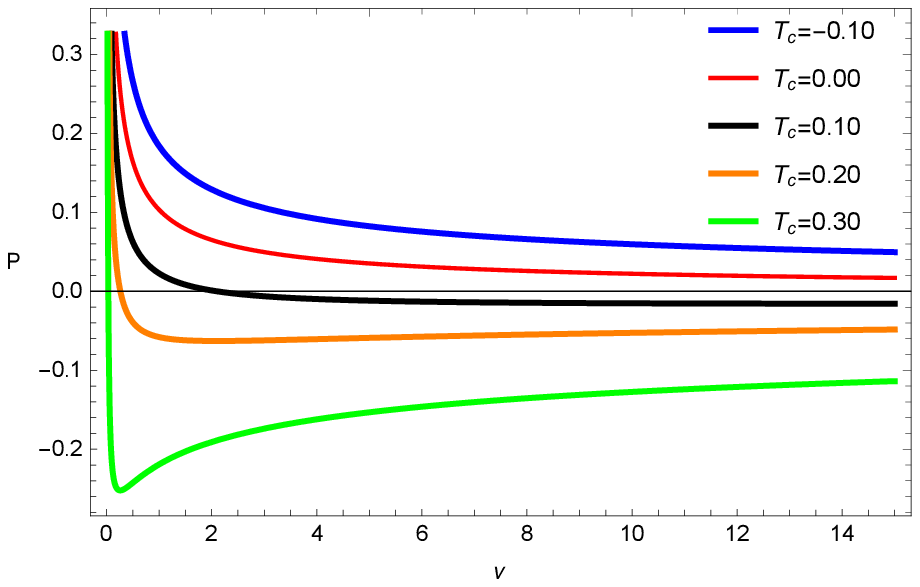}
\caption{\label{f3} Plot of temperature $P$ with fixed $q_m=0.0002$;
$d_1=0.004$; $f_1=0.003$; $\kappa _d =0.01$; $\kappa _s =0.03$ and $e
_1=0.4$.}
\end{minipage}\hspace{3pc}%
\begin{minipage}{14pc}
\includegraphics[width=16pc]{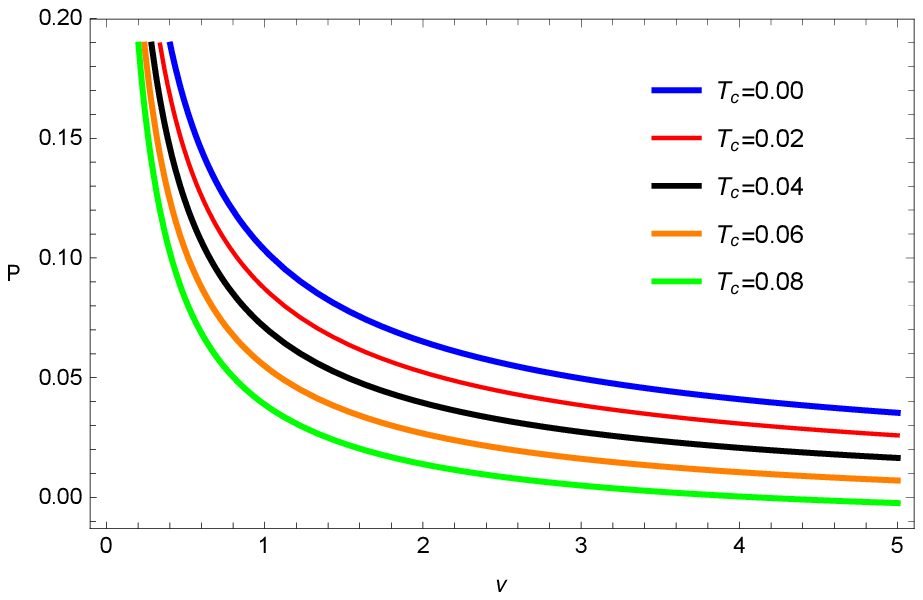}
\caption{\label{ff3} Plot of temperature $P$ with fixed $q_m=0.0002$;
$d_1=0.004$; $f_1=0.003$; $\kappa _d =0.01$; $\kappa _s =0.03$ and $e
_1=0.4$.}
\end{minipage}\hspace{3pc}%
\end{figure}

A cosmological constant is treated as a thermodynamic variable. After the thermodynamic pressure of the BH is putted into the laws of thermodynamics,
the cosmological constant is  considered as the pressure. From the equation of horizon $\Psi(r)=0$ and pressure $P=-\frac{\Lambda}{8\pi}$ \cite{26,27}, we can deduce the relation between the BH mass $m$ and its event horizon radius $r_h$, expression as follows
\begin{widetext}
\begin{equation}\label{T1}
m=\frac{3 d_1 \kappa_{s} ^2-6 f_1 \kappa_{\text{sh}}^2-8 \pi  P
r_h^4+3 q_e^2+3 q_m^2+3 r_h^2-12 \kappa_d ^2 e_{1} }{6 r_h}.
\end{equation}
\end{widetext}
The Hawking temperature of BH related to surface gravity can be
obtained  as
\begin{widetext}
\begin{equation}\label{T2}
T=\frac{\Psi'(r)}{4\pi}=\frac{6-32 \pi  P r_h^2}{12 \pi r_h}-\frac{3
d_1 \kappa_{s} ^2-6 f  \kappa_{\text{sh}}^2-8 \pi  P r_h^4+3 q_e^2+3
q_m^2+3 r_h^2-12 \kappa_d^2 e_{1} }{12 \pi  r_h^3}.
\end{equation}
\end{widetext}
It has a peak as shown in Figs. (\ref{f1}) and (\ref{f2}) and that
shifts to right (positive) and increases by increasing $P_c$ and
$\kappa_{s}$. The temperature  becomes the absence of the electric
charge  $(q= 0)$. As, we increase the values of $P_c$ and $\kappa_{s}$, the
the local maximum of the Hawking temperature increases in Figs.
(\ref{f1}) and (\ref{f2}). Further, the temperature converges when
the horizon radius shrinks to zero for the considered BH manifold.
The general form of the first law of BH thermodynamics can be written as \cite{26},\cite{27},\cite{28},\cite{29},\cite{t1},\cite{t2}
\begin{equation} \label{T3}
dM=TdS + V dP+ \Phi dq_m+  \varphi dq_e+\Bbbk_{sh} d\kappa_{sh}+\Bbbk_{s}d\kappa_{s}+\Bbbk_{d} d\kappa_{d}+E1 de_{1}+F_1 df_{1} +D_1 d d_{1},
\end{equation}
where $M$, $S$, $V$, $P$ , $Q$, $\Phi$ and $\varphi$ are the  mass, entropy, volume, pressure, magnetic charge, and chemical potential of BH, they have been treated as thethermodynamic variables corresponding to the conjugating variables $\Bbbk_{sh}$, $\Bbbk_{s}$, $\Bbbk_{d}$, $E1$ and $d_1$ respectively. The volume and chemical potential of BH are defined as
\begin{equation}\label{T4}
V=\bigg(\frac{\partial M}{\partial P}\bigg)_{S,q_m},
\Phi=\bigg(\frac{\partial M}{\partial q_m}\bigg)_{S,P},
\end{equation}
respectively. The BH entropy with the help of area is defined as \cite{t3, t4,yt1}
\begin{equation}\label{T5}
S=\frac{A}{4}=\pi r_{h}^{2}.
\end{equation}
From Eqs. (\ref{T1}) and (\ref{T2}), the equation of state for the BH can easily be expressed  as
\begin{equation}\label{T6}
P=-\frac{d_1 \kappa_{s}^2-2 f_1 \kappa_{\text{sh}}^2+q_e^2+q_m^2+4
\pi  r_h^3 T-r_h^2-4 \kappa_d ^2 e_{1} }{8 \pi  r_h^4}.
\end{equation}
Red, blue, orange, and black colors indicate the divergence at
pressures below the critical pressure.  The oscillations of the
isotherms at critical temperatures in the $P - v_h$ diagram are
equivalent to the unstable BHs that are presented by negative heat
capacity in this section (Figs. (\ref{f3}) and (\ref{ff3})). These
divergences are the characteristics of the first-order phase
transition that occurs between smaller and larger BHs that are
stable and have a positive heat capacity. In response to changes in
the value of the parameter $T_c$, there is a corresponding shift in
the horizontal axis; an increase in this parameter results in a
reduction in the critical radius.

The thermodynamic variables $V$, $\Phi$, $\varphi$ and the conjugating quantities $\Bbbk_{sh}$, $\Bbbk_{s}$, $\Bbbk_{d}$, $E1$ and $d_1$ are
obtained from the first law as
\begin{equation}\label{T7a}
V=\frac{4 \pi  r_h^3}{3}, \hspace{.1cm} \Phi=\frac{q_m}{r},\hspace{.16cm}  \varphi= \frac{q_e}{r}, \hspace{.16cm}  \Bbbk_{sh}= \frac{-2 f_1\kappa_{sh} }{r}, \hspace{.16cm} \Bbbk_{s}=\frac{ d_1\kappa_s }{r}, \hspace{.16cm}  \Bbbk_{d}=\frac{ -4 e_1\kappa_d}{r}, \hspace{.16cm}  E1= \frac{ -4 \kappa_d^2}{r}
\hspace{.1cm} \text{and} \hspace{.1cm}
D_1= \frac{ \kappa_s^2}{r}.
\end{equation}

\subsection{Gibbs Free Energy and Specific Heat}

The most important and basic thermodynamic quantity is Gibbs free of
BH, it can be utilized to explore small/larger BH phase transition
by studying $G-r_h$ and $G - T$ diagrams. In addition, Gibbs free
energy also helps us to investigate the global stability of BH. It
can be evaluated as \cite{t5,t6,t7}
\begin{equation}\label{T8}
G=-TS+M.
\end{equation}
Using Eqs.(\ref{T1}) and (\ref{T2}) in (\ref{T8}), we get
\begin{widetext}
\begin{equation}\label{T9}
G=\frac{\sqrt[3]{\frac{\pi }{6}} \left(12 d_1 \kappa_{s} ^2-24 f
\kappa_{\text{sh}} ^2+4 \sqrt[3]{\frac{6}{\pi }} P v^{4/3}+12
q_e^2+12 q_m^2+\left(\frac{6}{\pi }\right)^{2/3} v^{2/3}-48
\kappa_d^2 e_{1} \right)}{8 \sqrt[3]{v}}.
\end{equation}
\end{widetext}
\begin{figure}
\begin{minipage}{14pc}
\includegraphics[width=16pc]{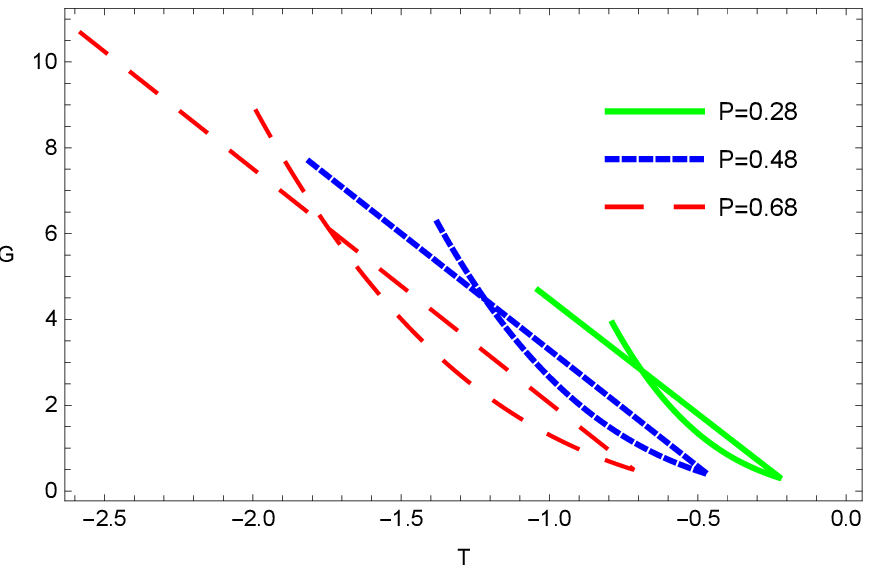}
\caption{\label{f5} Plot of Gibbs free energy $G$ with fixed
$q_m=0.003$; $d_1=0.200$; $f_1=0.050$; $\kappa _d =0.010$ and $e
_1=0.050$.}
\end{minipage}\hspace{3pc}%
\begin{minipage}{14pc}
\includegraphics[width=16pc]{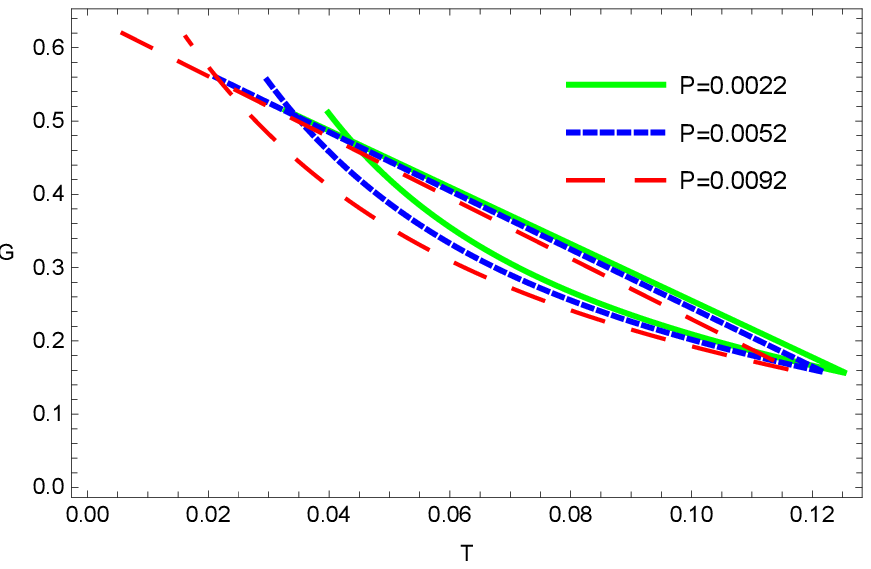}
\caption{\label{f6} Plot of Gibbs free energy $G$ with fixed
$q_m=0.003$; $d_1=0.200$; $f_1=0.050$; $\kappa _d =0.010$ and $e
_1=0.050$.}
\end{minipage}\hspace{3pc}%
\end{figure}
We observe the graphical behavior of the phase transitions in
$G-r_{h}$ plane as shown in Figs. (\ref{f5}) and (\ref{f6}). It is
noted that the Gibbs free energy decreases as the critical radius
increases.
To calculate the critical thermodynamic properties of BH one can use the following condition, given by \cite{t8}
\begin{equation}\label{T10}
\bigg(\frac{\partial P}{\partial
\upsilon_{h}}\bigg)_{T}=\bigg(\frac{\partial^{2} P}{\partial
\upsilon_{h}^{2}}\bigg)_{T}=0.
\end{equation}
Using Eq.(\ref{T10}), the critical temperature can be expressed as
\begin{equation}\label{T11}
T_{c}=\frac{1}{3 \sqrt{6} \pi \sqrt{d_1 \kappa_{s} ^2-2 f_1
\kappa_{\text{sh}} ^2+q_e^2+q_m^2-4 \kappa_d ^2 e_{1} }}.
\end{equation}
From Eq.(\ref{T10}), the critical radius of BH, yields as
\begin{equation}\label{T12}
\upsilon_{c}= 8 \sqrt{6} \pi  \left(d_1 \kappa_{s}^2-2 f_1
\kappa_{\text{sh}}^2+q_e^2+q_m^2-4 \kappa_d^2 e_{1} \right)^{3/2}.
\end{equation}
The critical pressure in terms of other parameters takes the
following form
\begin{widetext}
\begin{equation}\label{T13}
P_{c}=\frac{1}{96 \pi \left(-d_1 \kappa_{s}^2+2 f_1
\kappa_{\text{sh}}^2-q_e^2-q_m^2+4 \kappa_d^2 e_1 \right)^2}.
\end{equation}
\end{widetext}
However, we employed numerical analysis because calculating the critical numbers analytically is not a simple operation.

\begin{widetext}
\begin{figure}
\begin{minipage}{14pc}
\includegraphics[width=16pc]{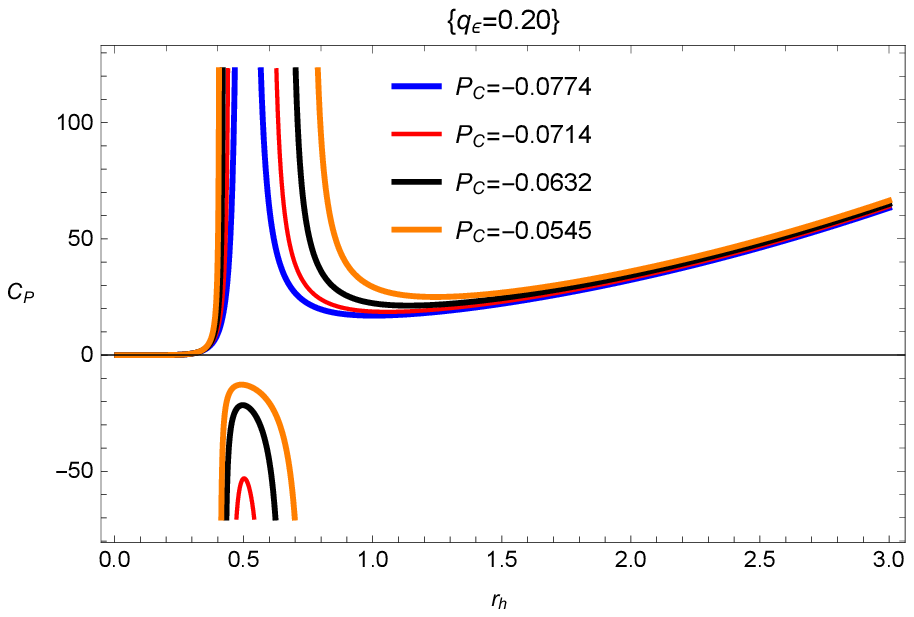}
\caption{\label{f7} Plot of Heat capacity with fixed  $q_m=0.08$;
$d_1=0.004$; $f_1=0.313$; $\kappa _d =0.02$ and $e _1=0.4$.}
\end{minipage}\hspace{3pc}%
\begin{minipage}{14pc}
\includegraphics[width=16pc]{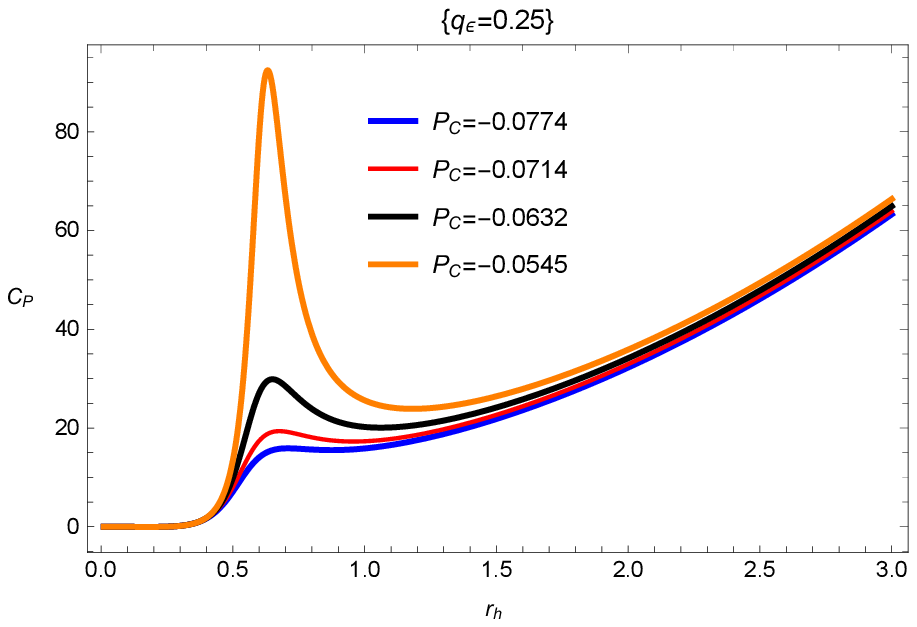}
\caption{\label{f8} Plot of Heat Capacity with fixed $q_m=0.08$;
$d_1=0.004$; $f_1=0.313$; $\kappa _d =0.02$ and $e _1=0.4$.}
\end{minipage}\hspace{3pc}%
\end{figure}
\end{widetext}

\begin{figure}
\begin{minipage}{14pc}
\includegraphics[width=16pc]{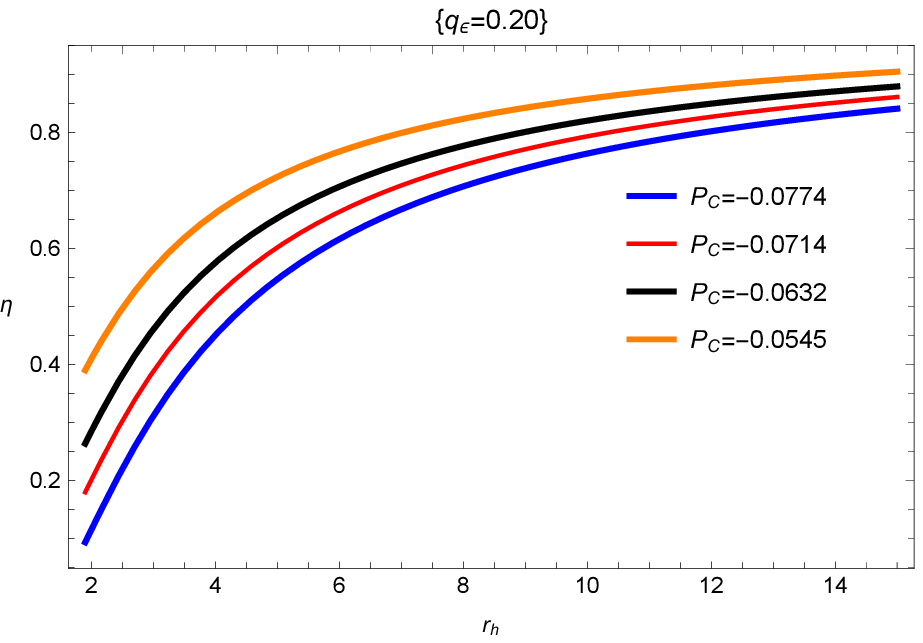}
\caption{\label{f9} Plot of Efficiency $\eta$ with fixed $q_m=0.08$;
$d_1=0.004$; $f_1=0.313$; $\kappa _d =0.02$ and $e _1=0.4$.}
\end{minipage}\hspace{3pc}%
\begin{minipage}{14pc}
\includegraphics[width=16pc]{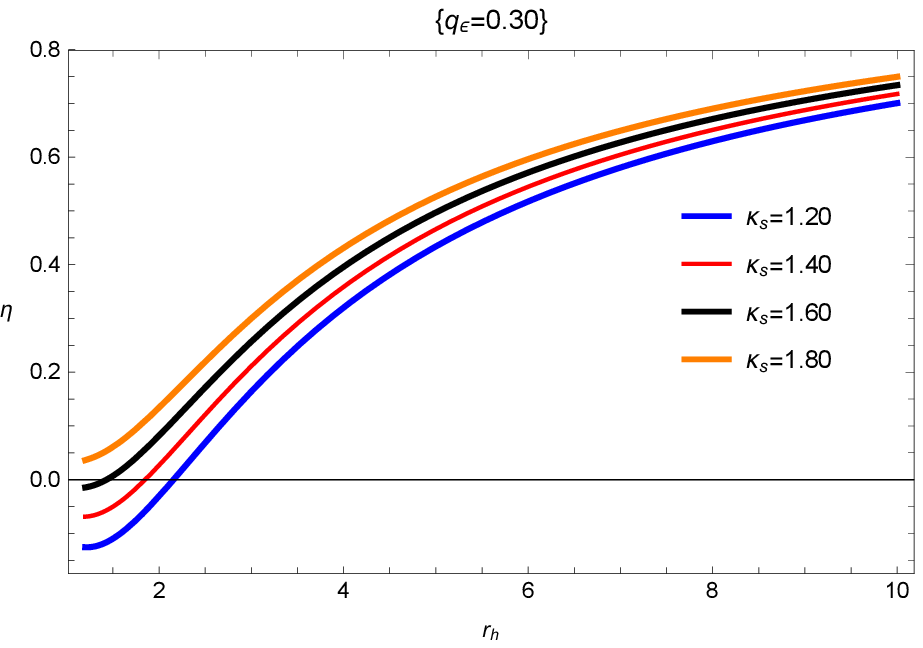}
\caption{\label{f10} Plot of Efficiency $\eta$ with fixed
$q_m=0.08$; $d_1=0.004$; $f_1=0.313$; $\kappa _d =0.02$ and $e
_1=0.4$.}
\end{minipage}\hspace{3pc}%
\end{figure}

%The critical points can be obtained numerically and present these points in Tab. \ref{tt1}.
%\begin{table}[htbp]
%\centering
%\begin{tabular}{|c|c|c|c|c|}
%\hline\hline
%$\kappa$  &$v_{c}$  &$T_{c}$  &$P_{c}$ \\
%\hline\hline
%$0.2$ &$0.522343$  & $0.212377$ &$-0.079705$ \\
%\hline
%$0.3$ &$0.545106$  & $0.209378$ &$-0.0774703$ \\
%\hline
%$0.4$ &$0.615308$  & $0.201092$ &$-0.0714597$  \\
%\hline
%$0.5$ &$0.738435$  & $0.189229$ &$-0.0632773$  \\
%\hline
%$0.6$ &$0.922934$  & $0.175672$ &$-0.054535$  \\
%\hline
%$0.7$ &$1.17946$  & $0.161881$ &$ -0.0463091$  \\
%\hline
%\end{tabular}
%\caption{\label{tt1} We obtained the critical values for setting
%$q_e=0.2$, $q_m=0.08$; $d_1=0.03$; $f_1=0.80$; $\kappa _d =0.05$ and
%$e _1=0.02$ under different values of $\kappa$.}
%\end{table}
To find  more data about a phase transition, we study thermodynamic a quantity such as heat capacity. By applying the  standard definition of heat capacity follows as \cite{t1, t9}
\begin{equation}\label{T14}
C_{p}=T \bigg(\frac{\partial S}{\partial T}\bigg)_{P},
\end{equation}
with little numerical calculations, one can get a dimensionless
important relation for the amounts $P_c$, $T_c$ and $v_c$. If
provided  expression $(d_1 \kappa_{s}^2-2 f_1
\kappa_{\text{sh}}^2+q_e^2+q_m^2-4 \kappa_d ^2 e_{1})\rightarrow
1.327765310$, then our solution satisfied the well-known condition
as
\begin{equation}\label{T140}
\frac{P_c v_c}{T_c} = 3/8,
\end{equation}
which is similar results are studied in the context of the Van der Waals equation and in RN-AdS BH \cite{t10}. Therefore, the negative heat capacity gives the temperamental (unstable) BH is also related to the critical temperature in $P-\upsilon_{h}$ plane. By using expressions of volume and entropy
of BH is studied in (\ref{T2}) and (\ref{T5}). From Eq.(\ref{T14}), we get
\begin{widetext}
\begin{equation}\label{T15}
C_{p}=\frac{3^{2/3} \sqrt[3]{\frac{\pi }{2}} v^{2/3} \left(4 d
\kappa_{s} ^2-8 f \kappa_{\text{sh}} ^2+12 \sqrt[3]{\frac{6}{\pi }}
P v^{4/3}+4 q_e^2+4 q_m^2-\left(\frac{6}{\pi }\right)^{2/3}
v^{2/3}-16 \kappa_d ^2 e_{1} \right)}{-12 d \kappa_{s} ^2+24 f
\kappa_{\text{sh}} ^2+12 \sqrt[3]{\frac{6}{\pi }} P v^{4/3}-12
q_e^2-12 q_m^2+\left(\frac{6}{\pi }\right)^{2/3} v^{2/3}+48 \kappa_d
^2 e_{1} }.
\end{equation}
\end{widetext}
It has been discovered that the critical amounts classify the
behavior of thermodynamic quantities close to the critical point. In
Figs. (\ref{f7}) and (\ref{f8}), For thermodynamically stable BHs,
we separate the two cases in which the heat capacity is positive ($r
_h <r_ c$) and the case in which it is negative ($r_ h> r_ c$). The
second-order phase transition is implied by the instability areas of
BHs, where the heat capacity is discontinuous at the critical
temperature $r _h=r _c$ \cite{t5a, t5b}. It is noted that the heat
capacity diverges at $r_h = 0.50$, when $T_h$ reaches its maximum
value as $T_h=0.24$ for $r_h = 1.00$, $q_m=0.08$, $d_1=0.004$,
$f_1=0.313$, $\kappa _d =0.02$ and $e _1=0.4$.
The critical points in alternate phase space are obtained by
utilizing the standard definition, we reduced the thermodynamic
variables as
\begin{widetext}
\begin{equation}\label{T16}
T_r=\frac{T}{T_c}, \quad v_r=\frac{v}{v_c}\quad and \quad
P_r=\frac{P}{P_c}.
\end{equation}
\end{widetext}
The reduced variables can be written as
\begin{widetext}
\begin{equation}\label{T17}
T_r=-\frac{3 \sqrt{\frac{3}{2}} \sqrt{d_1 \kappa_{s} ^2-2 f
\kappa_{\text{sh}} ^2+q_e^2+q_m^2-4 \kappa_d ^2 e_{1} } \left(d_1
\kappa_{s} ^2-2 f_1 \kappa_{\text{sh}} ^2+8 \pi  P
r_h^4+q_e^2+q_m^2-r_h^2-4 \kappa_d ^2 e_{1} \right)}{2 r_h^3},
\end{equation}
\end{widetext}
and volume can be obtained as
\begin{equation}\label{T18}
v_r=\frac{r^3}{6 \sqrt{6} \left(d_1 \kappa_{s} ^2-2 f_1
\kappa_{\text{sh}} ^2+q_e^2+q_m^2-4 \kappa_d ^2 e_{1}
\right)^{3/2}},
\end{equation}
and pressure follows as
\begin{widetext}
\begin{equation}\label{T19}
P_r=-\frac{12 \left(d_1 \kappa_{s} ^2-2 f_1 \kappa_{\text{sh}}
^2+q_e^2+q_m^2-4 \kappa_d ^2 e_{1} \right)^2 \left(d_1 \kappa_{s}
^2-2 f_1 \kappa_{\text{sh}} ^2+q_e^2+ q_m^2+4 \pi  r_h^3 T-r_h^2-4
\kappa_d^2 e_{1} \right)}{r_h^4}.
\end{equation}
\end{widetext}
Two adiabatic and two isothermal processes combine to form the
Carnot cycle is the hallmark of the most effective heat engine. The
single most fundamental and critical feature of the Carnot cycle is
that reservoir temperatures as a function of heat engine efficiency
\begin{widetext}
\begin{equation}\label{T20}
\eta= 1 -\frac {T_c}{T_h},
\end{equation}
\end{widetext}
As a reservoir can never be at zero temperature, the efficiency
cannot be one since $T_c$ is cold and $T_h$ denotes hot reservoirs.
Hence, we get
\begin{widetext}
\begin{equation}\label{T21}
\eta=1+ \frac{2 \sqrt{\frac{2}{3}} r_h^3}{3 \sqrt{d_1 \kappa_{s}
^2-2 f_1 \kappa_{\text{sh}} ^2+q_e^2+q_m^2-4 \kappa_d ^2 e_{1} }
\left(d \kappa_{s} ^2-2 f_1 \kappa_{\text{sh}} ^2+8 \pi  P
r_h^4+q_e^2+q_m^2-r_h^2-4 \kappa_d^2 e_{1} \right)}.
\end{equation}
\end{widetext}
Now, we study the behavior of the heat engine efficiency $\eta$ as a
function of the the pressure $P$ and entropy $S$ matching to the
heat cycle provided in Figs. (\ref{f9}) and (\ref{f10}), for the
different values of metric-affine gravity parameters. From these
figures, we distinguish that the nature of the heat engine
efficiency is essentially relying on the metric-affine gravity
parameters. In addition, for a given set of input values, the
efficiency of the heat engine increases monotonically as the
horizon's radius grows. Because of this, larger BHs should expect
higher heat-engine efficiency. In other words, they allow for a
maximum efficiency curve to be provided for a heat engine by varying
only a few fixed parameters (BH works at the highest efficiency).
Here, local stability is related to the system, but it can be great
to the small changes in the values of thermodynamic parameters.
Thus, the term heat capacity gives information on local stability.
In \cite{y5}, it is stated how the cosmological constant $\Lambda$
can be studied by treating it as a scale parameter.

\section{Joule-Thomson Expansion}
\begin{figure}
\begin{minipage}{14pc}
\includegraphics[width=16pc]{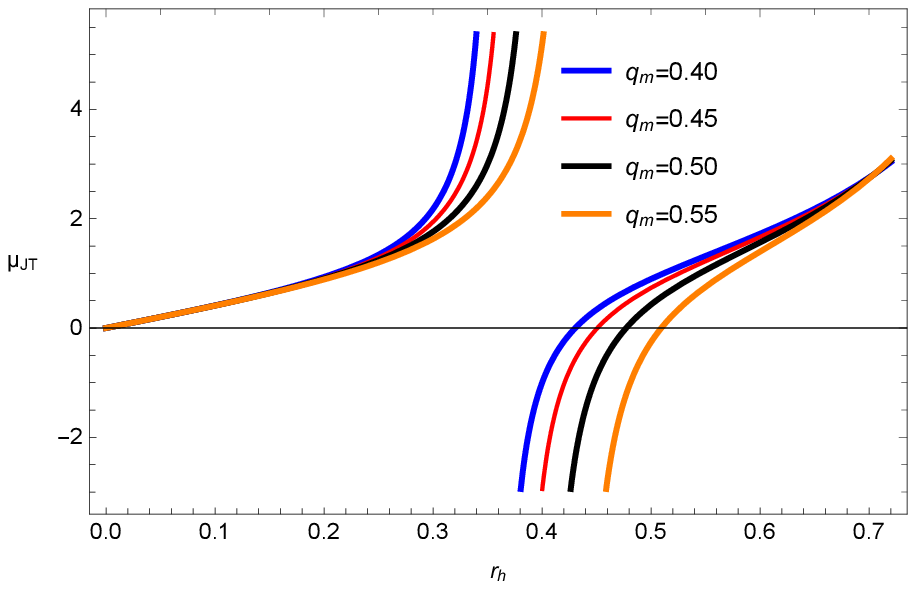}
\caption{\label{f19} Joule-Thomson coefficient $\mu_{JT}$ Plane with
fixed $d_1=0.03$; $f_1=0.01$; $\kappa _{d} =0.02$;
$\kappa _s =0.10$ and $e _1=0.04$.}
\end{minipage}\hspace{3pc}%
\begin{minipage}{14pc}
\includegraphics[width=16pc]{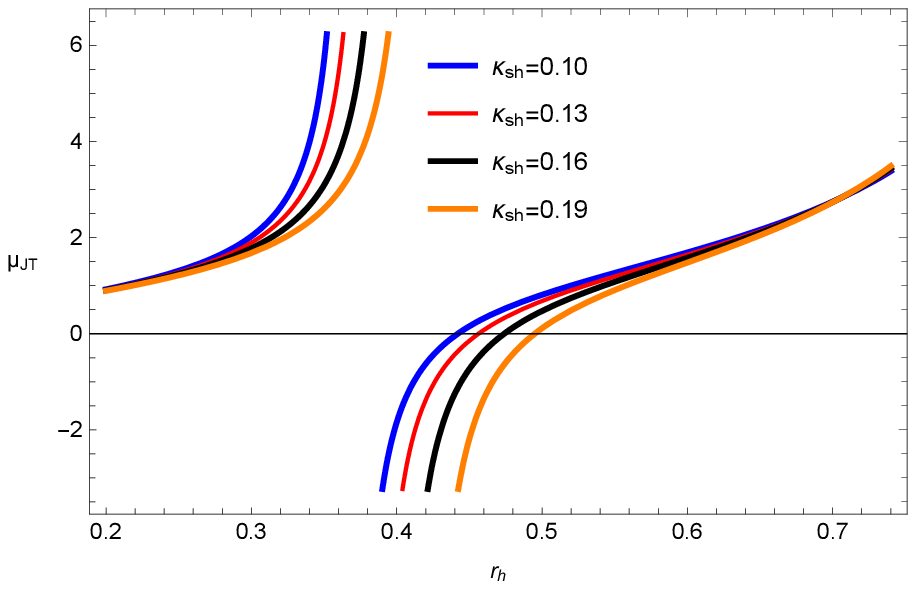}
\caption{\label{f20} Joule-Thomson coefficient $\mu_{JT}$ with fixed
$d_1=0.03$; $f_1=0.01$; $\kappa _{d} =0.02$;
$\kappa _s =0.10$ and $e _1=0.04$.}
\end{minipage}\hspace{3pc}%
\begin{minipage}{14pc}
\includegraphics[width=16pc]{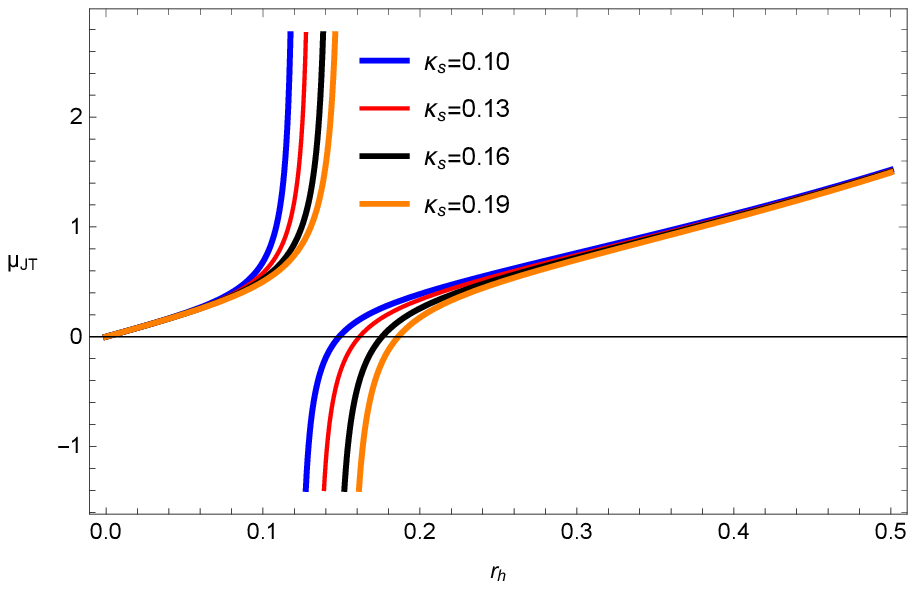}
\caption{\label{f21} Joule-Thomson coefficient $\mu_{JT}$ with fixed
$d_1=0.03$; $f_1=0.01$; $\kappa _{sh} =0.10$;
$\kappa _d =0.02$ and $e _1=0.04$.}
\end{minipage}\hspace{3pc}%
\begin{minipage}{14pc}
\includegraphics[width=16pc]{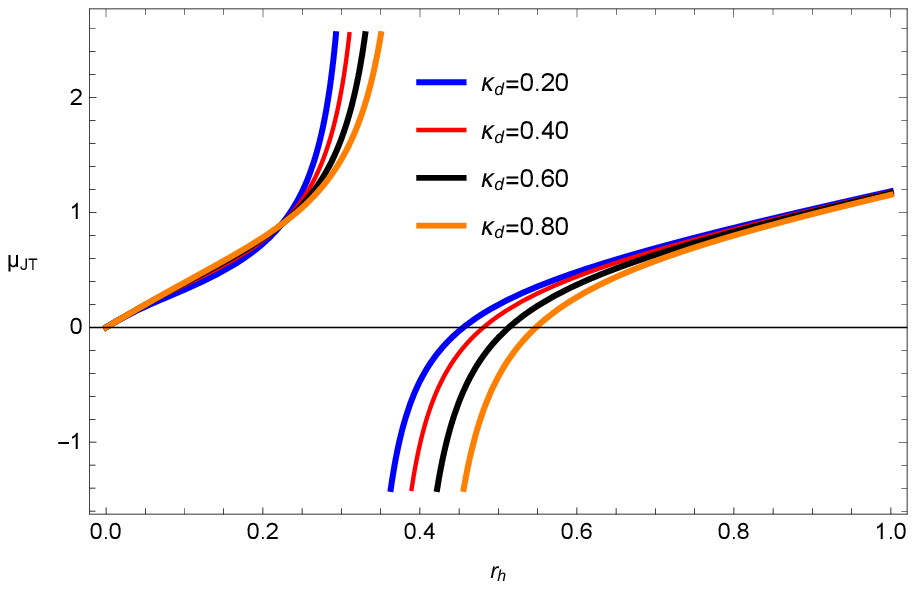}
\caption{\label{f22} Joule-Thomson coefficient $\mu_{JT}$ with fixed
$d_1=0.03$; $f_1=0.01$; $\kappa _{sh} =0.10$;
$\kappa _s =0.10$ and $e _1=0.04$.}
\end{minipage}\hspace{3pc}
\end{figure}

\begin{figure}
\begin{minipage}{14pc}
\includegraphics[width=16pc]{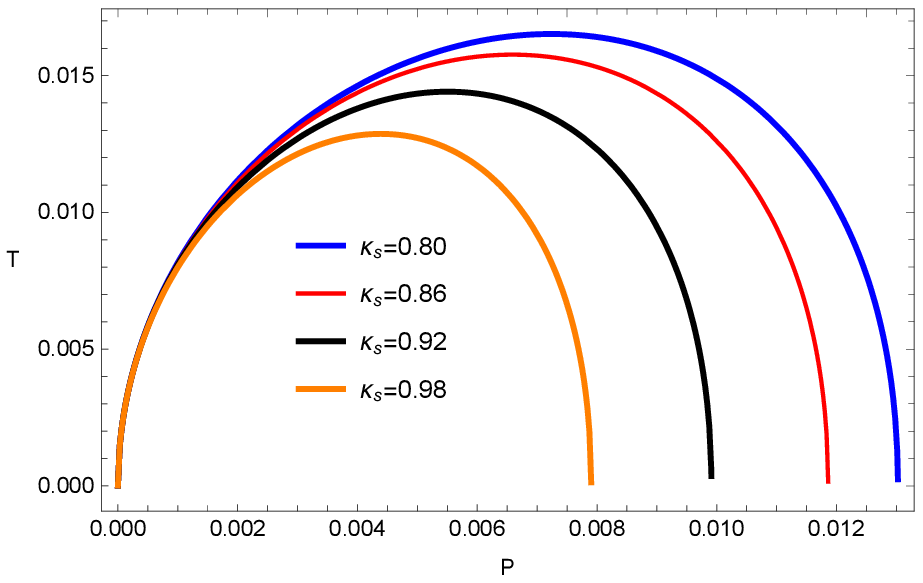}
\caption{\label{f15} Isenthalpic curves $(T-P)$ Plane with fixed
$d_1=0.004$; $f_1=0.313$; $\kappa _{sh} =0.05$; $\kappa _d =0.02$;
and $e _1=0.4$.}
\end{minipage}\hspace{3pc}%
\begin{minipage}{14pc}
\includegraphics[width=16pc]{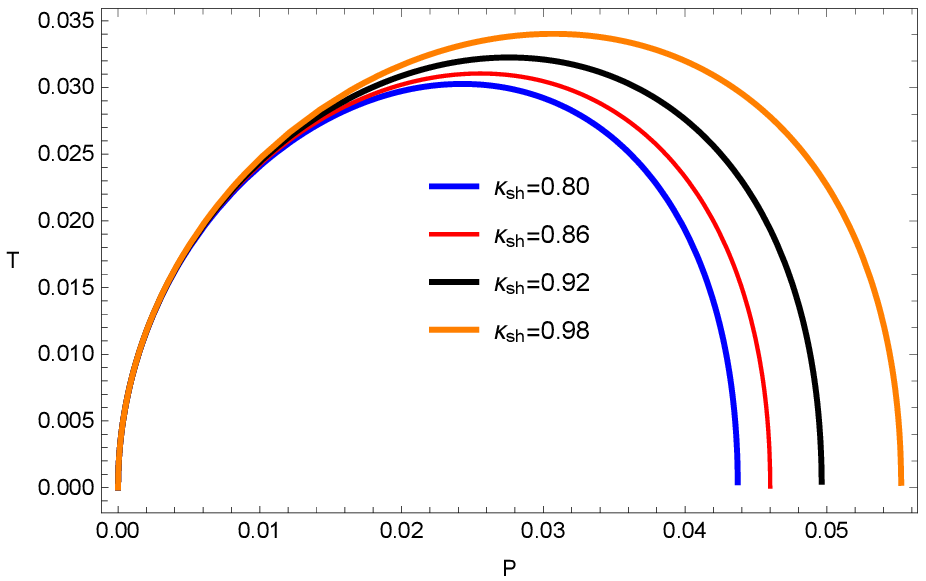}
\caption{\label{f16} Isenthalpic curves $(T-P)$ Plane with fixed
$d_1=0.004$; $f_1=0.313$; $\kappa _d =0.02$; $\kappa _s =0.8$ and $e
_1=0.4$.}
\end{minipage}\hspace{3pc}%
\begin{minipage}{14pc}
\includegraphics[width=16pc]{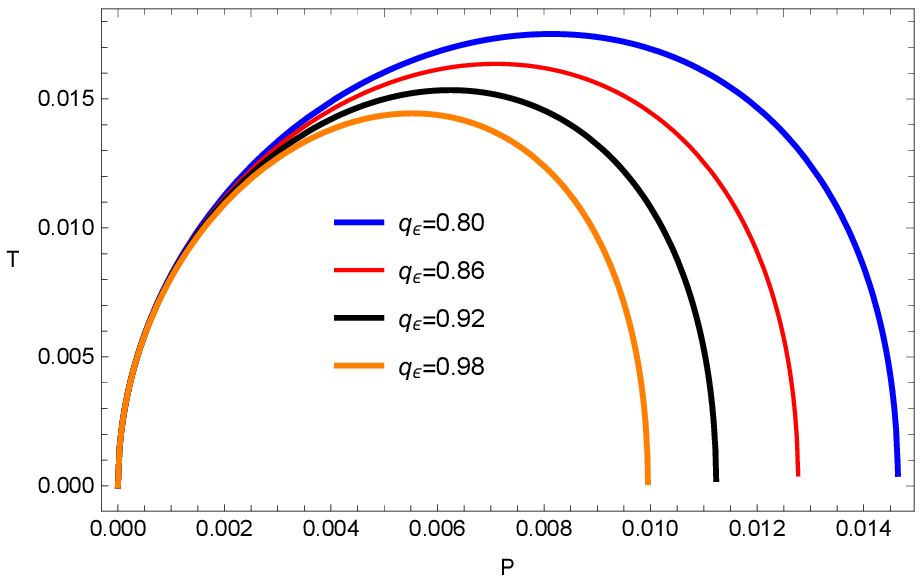}
\caption{\label{f17} Isenthalpic curves $(T-P)$ Plane with fixed
$d_1=0.004$; $f_1=0.313$; $\kappa _{sh} =0.05$; $\kappa _d =0.02$;
$\kappa _s =0.8$ and $e _1=0.4$.}
\end{minipage}\hspace{3pc}%
\begin{minipage}{14pc}
\includegraphics[width=16pc]{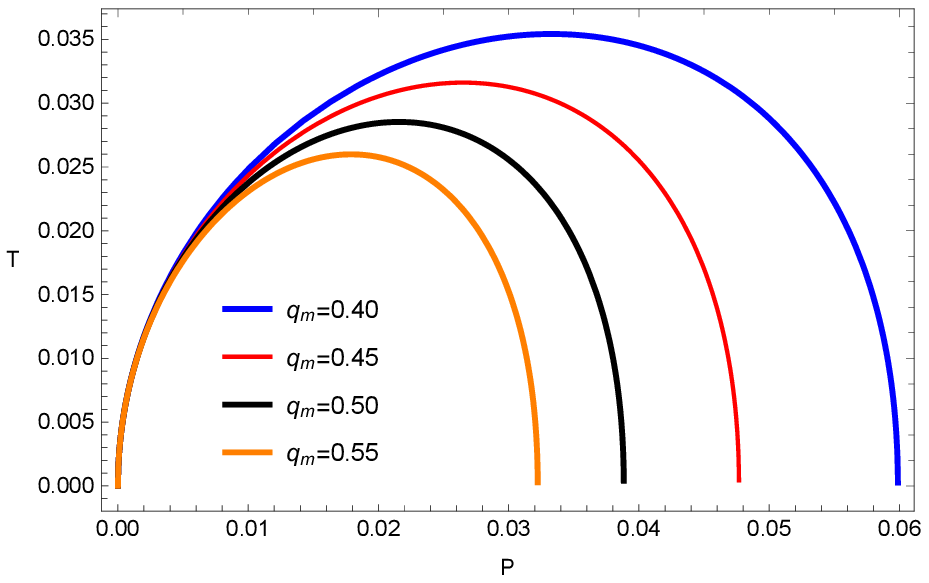}
\caption{\label{f18} Isenthalpic curves ${T-P}$ Plane with fixed
$d_1=0.004$; $f_1=0.313$; $\kappa _{sh} =0.05$; $\kappa _d =0.02$;
$\kappa _s =0.8$ and $e _1=0.4$.}
\end{minipage}\hspace{3pc}%
\end{figure}
One of the most well-known and classical physical process to explain the change in the temperature of gas from a high-pressure section to reduced pressure
through a porous plug is  called  Joule-Thomson expansion. The main focus is on the gas expansion process, which expresses the cold effect (when the temperature drops) and the heat effect (when the temperature increases), with the enthalpy remaining constant throughout the process. This change depends upon the coefficient of Joule-Thomson as \cite{22a,j2}
\begin{widetext}
\begin{equation}\label{j1}
\mu_{JT}=\bigg(\frac{\partial T}{\partial
P}\bigg)_{H}=\frac{1}{C_{p}}\bigg[T\bigg(\frac{\partial V}{\partial
T}\bigg)_{p}-V\bigg].
\end{equation}
\end{widetext}
Using  Eqs.(\ref{T2}), (\ref{T4}), (\ref{T15}) and (\ref{j1})
coefficient calculated as
\begin{widetext}
\begin{equation} \label{j2}
\mu_{JT}=\frac{4 r_h \left(3 d_1 \kappa_{s} ^2-6 f_1
\kappa_{\text{sh}} ^2+8 \pi  P r_h^4+3 q_e^2+3 q_m^2-2 r_h^2-12
\kappa_d ^2 e_{1} \right)}{3 \left(d_1 \kappa_{s} ^2-2 f
\kappa_{\text{sh}} ^2+8 \pi  P r_h^4+q_e^2+q_m^2-r_h^2-4 \kappa_d ^2
e_{1} \right)}.
\end{equation}
\end{widetext}
The study of coefficient of Joule-Thomson versus the horizon $r_h$
is shown in Figs. (\ref{f19}), (\ref{f20}), (\ref{f21}) and
(\ref{f22}). We set $d_1=0.004$, $f_1=0.313$, $\kappa _{sh} =0.05$,
$\kappa _d =0.02$, $\kappa _s =0.8$ and $e _1=0.4$ in the order.
There exist both divergence points and zero points for different
variations $\kappa _d $, $\kappa _s$, and $\kappa _{sh}$ respectively. It is
clear from a comparison of these figures that the zero point of the
Hawking temperature and the divergence point of the coefficient of
Joule-Thomson is the same. This point of divergence gives
information on the Hawking temperature and corresponds to the most
extreme BHs. From Eq.(\ref{j2}) utilizing the well known condition $\mu_{JT}=0$, the temperature inversion occurs as
\begin{widetext}
\begin{equation} \label{j3}
T_i=\frac{3 d_1 \kappa_{s} ^2-6 f_1 \kappa_{\text{sh}} ^2-8 \pi  P
r_h^4+3 q_e^2+3 q_m^2-r_h^2-12 \kappa_d ^2 e_{1}}{12 \pi r_h^3}.
\end{equation}
\end{widetext}
Since the Joule-Thomson expansion is an isenthalpic process, it is
important to analyze the isenthalpic curves of BHs under
metric-affine gravity that is depicted in Figs.
(\ref{f15})-(\ref{f18}). So, we study isenthalpic curves ($T_i
-P_i$- plane) by assuming different values of BH mass which investigated in  Eq.(\ref{j2}) with a larger root of $r_h$.
We show the isenthalpic curves and the inversion curves of BH in
metric-affine gravity and this result is consistent
\cite{j3,j4,j5,j6}. Heating and cooling zones are characterized by
the inversion curve, and isenthalpic curves possess positive slopes
above the inversion curve. In contrast, the pressure always falls in
a Joule-Thomson expansion and the slope changes sign when heating
happens below the inversion curve. The heating process appears at
higher temperatures, as indicated by the negative slope of the
constant mass curves in the Joule-Thomson expansion. When
temperatures drop, cooling begins, which is linked to the positive
slope of the constant mass curves.
\begin{figure}
\begin{minipage}{14pc}
\includegraphics[width=16pc]{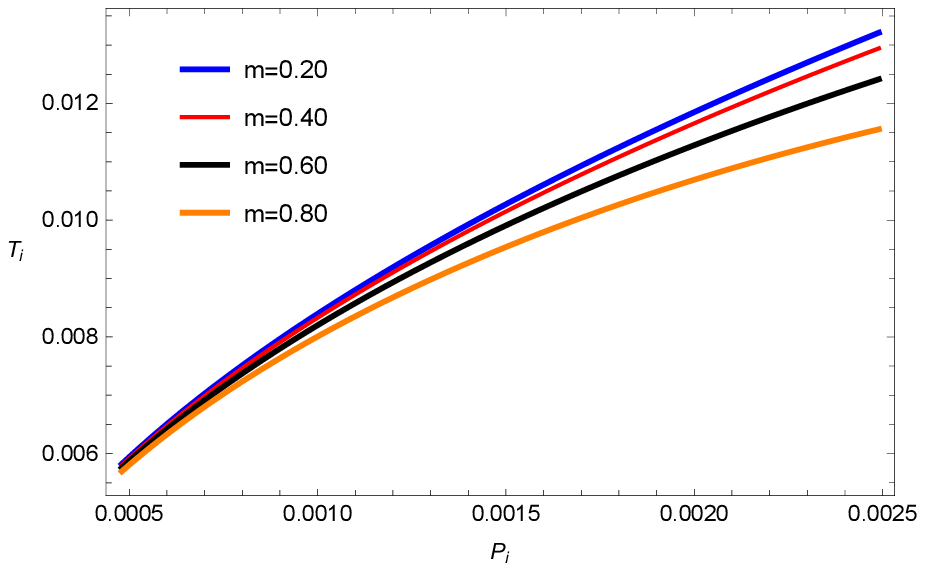}
\caption{\label{f11} Inversion curves $(Ti-Pi)$  with fixed
$d_1=0.004$; $f_1=0.313$; $\kappa _{sh} =0.05$; $\kappa _d =0.02$;
$\kappa _s =0.8$ and $e _1=0.4$.}
\end{minipage}\hspace{3pc}%
\begin{minipage}{14pc}
\includegraphics[width=16pc]{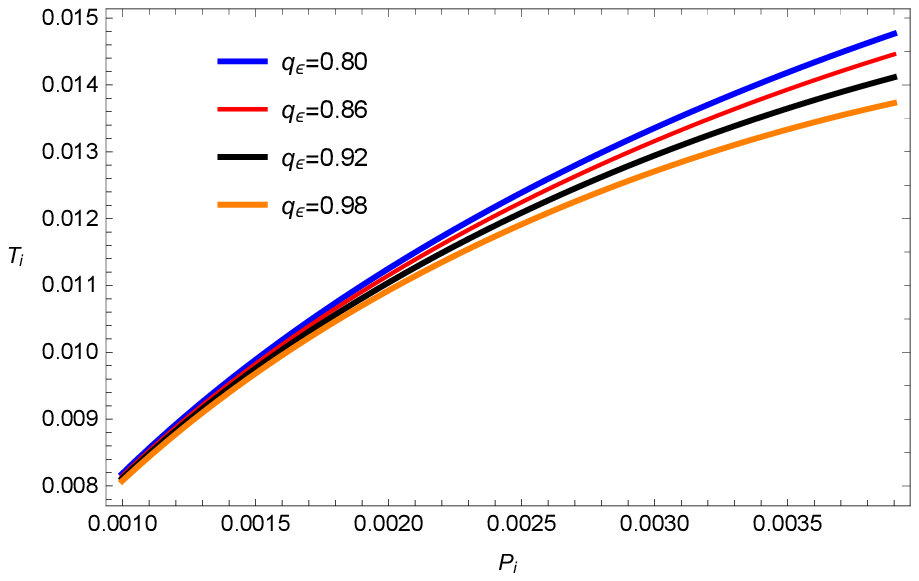}
\caption{\label{f12} Inversion curves $(Ti-Pi)$  with fixed
$d_1=0.004$; $f_1=0.313$; $\kappa _{sh} =0.05$; $\kappa _d =0.02$;
$\kappa _s =0.8$ and  $e _1=0.4$.}
\end{minipage}\hspace{3pc}%
\begin{minipage}{14pc}
\includegraphics[width=16pc]{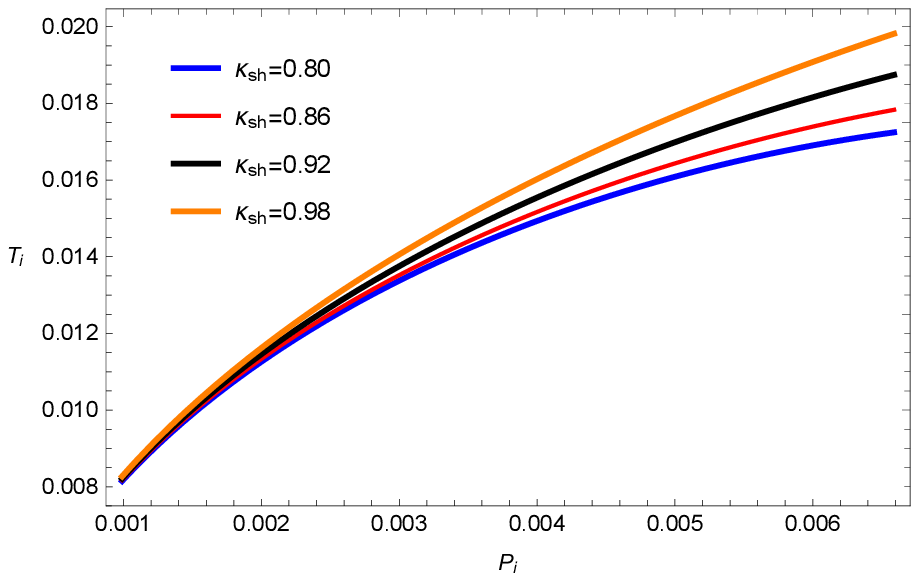}
\caption{\label{f13} Inversion curves $(Ti-Pi)$ Plane with fixed
$d_1=0.004$; $f_1=0.313$; $\kappa _d =0.02$; $\kappa _s =0.8$ and $e
_1=0.4$.}
\end{minipage}\hspace{3pc}%
\begin{minipage}{14pc}
\includegraphics[width=16pc]{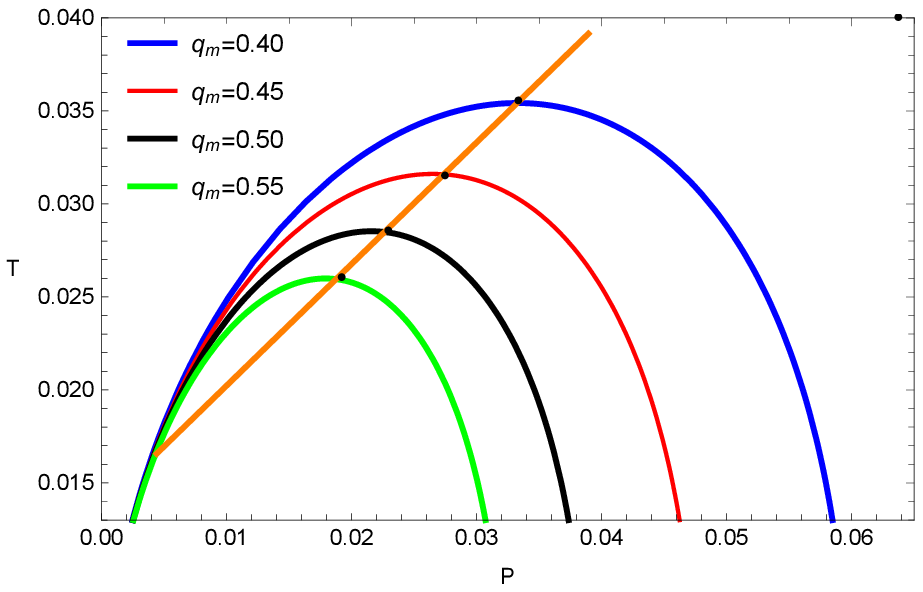}
\caption{\label{f14} Inversion curves $(Ti-Pi)$ Plane with fixed
$d_1=0.004$; $f_1=0.313$; $\kappa _{sh} =0.05$; $\kappa _d =0.02$
and  $e _1=0.4$.}
\end{minipage}\hspace{3pc}%
\end{figure}
From above equation, one can deduced the inversion pressure as
\begin{widetext}
\begin{equation} \label{j4}
P_i =\frac{3 d_1 \kappa_{s} ^2-6 f_1 \kappa_{\text{sh}} ^2+3
q_e^2+3q_m^2-12 \pi  r_h^3 Ti-r_h^2-12 \kappa_d^2 e_{1} }{8 \pi
r_h^4}.
\end{equation}
\end{widetext}
The inversion curves for different values  $d_1=0.004$, $f_1=0.313$,
$\kappa _{sh} =0.05$, $\kappa _d =0.02$ are shown in Figs.
(\ref{f11}), (\ref{f12}), (\ref{f13}) and (\ref{f14}). The inversion
temperature increases with variations of important parameters  $m$,
$q_e$, $\kappa _{sh}$ and $\kappa _s$ respectively. We can go back to the case of
BH in metric-affine gravity. Compared with the van der Waals fluids,
we see from Figs. (\ref{f11})-(\ref{f14}) that the inversion curve
is not closed. From the above results, in $Ti-Pi$-plan at low pressure, the inversion temperature $T_i$ decreases with the increase of charge $q_e$ and mass $m$, which shows the opposite behavior for higher pressure. It is also clear that, unlike the case with van der Waals fluids, the inversion temperature continues to rise monotonically with increasing inversion pressure, and hence inversion curves are not closed \cite{j5,j6}.

\section{Conclusion}

In this paper, we have considered BH in metric-affine gravity and
studied thermodynamics in presence of Bekinstien entropy, and
examined the standard thermodynamics relations. In detail, we have
thoroughly investigated thermodynamics to analytically obtain
thermodynamical properties like the Hawking temperature, entropy,
specific heat, and free energy associated with BH in metric-affine
gravity with a focus on the stability of the system. The heat
capacity blows at $r_c$, which is a double horizon, and local maxima
of the Hawking temperature also occur at $r_c$. It is shown that the
heat capacity is positive for $r_h < r_c$ providing the stability of
small BHs close-up to perturbations in the region, and at critical
radius phase transition exists. While the BH is unstable for $r_h >
r_c$ with negative heat capacity. The global analysis of the
stability of BH is also discussed by calculating free energy $G_h$.
For negative free energy $G_h < 0$ and positive heat capacity $C_p >
0$, it is noted that smaller BHs are globally stable, and also these
results are used in Refs. \cite{15,16,17,18,19,20}. We calculated the inverse temperature, inverse pressure and mass parameter, which investigated the  Joule-Thomson process of the system. The negative cosmological constant in metric-affine gravity is investigated to phase transitions of BHs. Above the inversion curves, we examined  the cooling region, while below the inversion  curve it leads to the heating one. The corresponding results can be summarized as: both the inversion temperature and pressure are become greater with the increasing of BH in metric-affine gravity while they are decreasing with the charge $q_e$. The physical consequences was analogous with the holography, where BHs would being a system as well as  dual to conformal field theories. The BH in metric-affine gravity is studied and  identical to the thermodynamics of usual systems and their thermodynamical analysis become more complete. Our results show a characteristic of Joule-Thomson coefficient is independent of the shear, spin and dilation charges indicating that the Joule-Thomson expansion we consider here is universal. In particular, we find a novel isenthalpic curves  in which the inversion temperature of the Joule-Thomson expansion, rather than the extreme one reported by previous work separates the analogues to heating-cooling phase \cite{H1,H2,H3}. Therefore, our inversion curves separates the allowable along with  forbidden regions for the Joule-Thomson effect to be observed, where the Joule-Thomson coefficient is the essential quantity to discriminate between the cooling and heating regimes of the system. It is worth noting that when amplify  a thermal system with a temperature so that pressure always decreases yielding a negative sign to $\partial P$.
 
Our analysis of inversion curves in the plane revealed that the influence of the  parameters a BH may be more evident in space-time.  We analyzed the BH in metric-affine gravity characteristics on the inversion curve; these included the the shear, spin and dilation charges. In these figures, we observed  that the inversion curves be compatible with the extreme point of a specific isenthalpic curves, and the cooling as well as heating regions are identified. In other words, the boundary between the heating-cooling regions of the BH in  metric-affine gravity influence on the inversion curves. We also discovered that both the maximum expansion points of cooling-heating regimes like \cite{H4,H5,H6}. For the BH heat engine, we investigated the analytical expression for the efficiency
in terms of horizon radius, pressures and temperatures in various
limits. We have also studied the Joule-Thomson expansion,
isenthalpic curves and inversion curves of the considered BH in
metric affine gravity as given below:
\begin{itemize}
\item We have examined the Joule-Thomson expansion for BH
in metric-affine gravity, where the cosmological constant is taken
as a pressure. We mainly focused on BH mass is considered enthalpy,
it is the mass that does not change during the expansion. The
Joule-Thomson coefficient $\mu_{JT}$ in terms of  horizon $r_h$ is
shown in Figs. (\ref{f19}), (\ref{f20}), (\ref{f21}) and
(\ref{f22}). There exist both divergence points and zero points with
$d_1=0.004$, $f_1=0.313$, $\kappa _{sh} =0.05$, $\kappa _d =0.02$,
$\kappa _s =0.8$ and $e _1=0.4$. The zero point of the Hawking
temperature, which is related to the most distant BHs, agrees with
the divergence point of the Joule-Thomson coefficient, which is
depicted in a consistent manner \cite{j5,j6}.

\item We also presented the isenthalpic curves such results are
presented in higher dimensions as demonstrated in Figs. (\ref{f15}),
(\ref{f16}), (\ref{f17}) and (\ref{f18}). It is very interesting to
explain that the positive slopes of the inversion curve are found as
mentioned in the literature \cite{j2,22a}. This indicates that with
the expansion of a metric-affine universe, BH always cools above the
inversion curve.

\item To determine the temperature gradients between the cooling and heating zones for various values of $d_1$, $f_1$, $\kappa _{sh}$
and $\kappa _d$, we have analyzed the inversion curve (Figs.
(\ref{f11})-(\ref{f14})).
\end{itemize}
It is concluded that the considered BH in metric affine gravity
meets the results in the literature and also this work is beneficial
for future research.

\section*{Acknowledgement}
This project was supported by the natural sciences foundation of
China (Grant No. 11975145). Faisal Javed acknowledges Grant No.
YS304023917 to support his Postdoctoral Fellowship at Zhejiang
Normal University, China. 
\section*{Declaration of competing interest}
The authors declare that they have no known competing financial interests or personal relationships that could have appeared to
influence the work reported in this paper.

\section*{Data Availability Statement} This manuscript has no associated data, or the data will not be deposited.
(There is no observational data related to this article. The
necessary calculations and graphic discussion can be made available
on request.)

\end{document}